\documentclass[12pt]{iopart}
\usepackage{amsbsy,amssymb,amstext,rsfs}  
\newcommand{\be}{\begin{equation}}
\newcommand{\ee}{\end{equation}}
\newcommand{\ba}{\begin{array}}
\newcommand{\ea}{\end{array}}
\newcommand{\bea}{\begin{eqnarray}}
\newcommand{\eea}{\end{eqnarray}}
\newcommand{\bean}{\begin{eqnarray*}}
\newcommand{\eean}{\end{eqnarray*}}
\newcommand{\lp}{\left(}
\newcommand{\rp}{\right)}
\newcommand{\ls}{\left[}
\newcommand{\rs}{\right]}
\newcommand{\lc}{\left\{}
\newcommand{\rc}{\right\}}

\newcommand{\la}{\langle}
\newcommand{\La}{\left\la}
\newcommand{\ra}{\rangle}
\newcommand{\Ra}{\right\ra}

\newcommand{\gb}%
	{\text{\bf\slshape g%
	}}

\newcommand{\intl}{\int\limits}
\newcommand{\im}{{\rm i}}
\newcommand{\veps}{\varepsilon}
\newcommand{\vphi}{\varphi}
\newcommand{\vrho}{\varrho}
\newcommand{\mPsi}{{\mit\Psi}}
\newcommand{\SP}{{\scr P}}
\newcommand{\Sp}{\mathop{\rm Sp}\nolimits}

\newcommand{\ddt}{\frac{\partial}{\partial t}}
\newcommand{\refp}[1]{(\ref{#1})}
\newcommand{\ds}{\displaystyle}
\newcommand{\bmv}[1]{\boldsymbol #1}
\newcommand{\av}{a^{\dag}}
\newcommand{\ai}{a^{\phantom{\dag}}}
\newcommand{\hav}{\hat{a}^{\dag}}
\newcommand{\tav}{\tilde{a}^{\dag}}
\newcommand{\hai}{\hat{a}^{\phantom{\dag}}}
\newcommand{\tai}{\tilde{a}^{\phantom{\dag}}}
\newcommand{\hgv}{\hat{\gamma}^{\dag}}
\newcommand{\tgv}{\tilde{\gamma}^{\dag}}
\newcommand{\hgi}{\hat{\gamma}^{\phantom{\dag}}}
\newcommand{\tgi}{\tilde{\gamma}^{\phantom{\dag}}}
\newcommand{\php}{{\phantom B}}
\newcommand{\vrhoq}{\vrho_{\rm q}}
\newcommand{\hvrhoq}{\hat{\vrho}_{\rm q}}
\newcommand{\vt}{\vartriangle\!\!}
\renewcommand{\d}{\mbox{\rm d}}

\begin{document}
\jl{4}
\eqnobysec

\title[Kinetics and hydrodynamics of dense quantum systems]
{Transport equations of a consistent description of the 
kinetics and hydrodynamics of dense quantum systems.\\
I: General approach within the frame of nonequilibrium thermo field dynamics}

\author{M V Tokarchuk and A E Kobryn}

\address{Institute for Condensed Matter Physics,\\
1~Svientsitskii St., UA--290011 Lviv, Ukraine}

\begin{abstract}
We present basic equations of nonequilibrium thermo field dynamics of dense 
quantum systems. A formulation of nonequilibrium thermo field dynamics has 
been performed using the nonequilibrium statistical operator method by 
D.N.Zubarev. Generalized transfer and hydrodynamic equations of a consistent 
description of kinetics and hydrodynamics have been obtained in thermo field 
representation. To demonstrate how obtained results do work at the 
description of kinetics and hydrodynamics of a dense nuclear matter we 
consider quantum system with strongly coupled states.
\end{abstract}


%05.30.-d quantum statistical mechanics
%05.60.+w transport processes, theory

\submitted
\section{Introduction}

The development of methods for the construction of kinetic and hydrodynamic 
equations in the theory of nonequilibrium processes for temperature quantum
field systems is, in particular, important for the investigation of 
nonequilibrium properties of a quark-gluon plasma \cite{c1,c2,c3,c4,c5} -- 
one of the nuclear matter states which can be created at ultrarelativistic 
collisions of heavy nuclei \cite{c6,c7,c8,c9}. In the studies of 
nonequilibrium states of quantum field systems, such as a nuclear matter 
\cite{c8,c9}, there arises a problem of taking into consideration coupled 
states. Kinetic and hydrodynamic processes in a hot, compressed nuclear 
matter, which appears after ultrarelativistic collisions of heavy nuclei or 
laser thermonuclear synthesis, are mutually connected and we should 
consider coupled states between nuclons. This is of great importance for 
the analysis and correlation of final reaction products. Obviously, a 
nuclon interaction investigation based on a quark-gluon plasma is a 
sequential microscopic approach to the dynamical description of reactions 
in a nuclear matter. For the description of kinetic processes in a nuclear 
matter on the level of model interactions, the Vlasov-Uehling-Uhlenbeck 
kinetic equation is used. This equation is used mainly in the case of low 
densities. The problems of a dense quark-gluon matter were discussed in 
detail in \cite{c8,c9,c20,c45,c46,c47}. As this takes place, its density 
increases by a factor of ten in the fourth degree and the distance between 
nuclons in the centre reaches $\sim 10^{-13}$ cm. Such systems are examples 
of strong both long-range and short-range (nuclear) interactions. There is 
no small parameter for these systems (density, for example). Nonequilibrium 
processes have a strongly correlated collective nature. That is why methods 
which are based on a one-particle description, in particular, on the basis 
of Boltzmann-like kinetic equations, cannot be used. In addition to high 
temperature dense quantum systems, there are Bose and Fermi systems at low 
temperatures with decisive many-particle dissipative correlations. Neither 
the linear response theory nor Boltzmann-like kinetic equations are 
sufficient for their description.

Analysis of the problem of a description of kinetic processes in highly 
non\-equi\-li\-bri\-um and strongly coupled quantum systems on the basis of 
the nonequilibrium real-time Green functions technique \cite{c14,c15,c16} 
and the theory in terms of non-Markovian kinetic equations describing memory
effects \cite{c17,c18,c19} was made in recent paper \cite{c20} and then in 
monograph \cite{c21}. It is important to note that in \cite{c20} the quantum 
kinetic equation for a dense and strongly coupled nonequilibrium system was 
obtained when the parameters of a shortened description included a 
one-particle Wigner distribution function and an average energy density. On 
the basis of this approach the quantum Enskog kinetic equation was obtained 
in \cite{c21}. This equation is the quantum analogue of the classical one 
within the revised Enskog theory \cite{c22,c23}. Problems of the 
construction of kinetic and hydrodynamic equations for highly 
nonequilibrium and strongly coupled quantum systems were considered based 
on the nonequilibrium thermo field dynamics in 
\cite{c24,c25,c26,c27,c28,c29}. In particular, a generalized kinetic 
equation for the average value of the Klimontovich operator was obtained in 
\cite{c25} with the help of the Kawasaki-Gunton projection operator method 
\cite{c30}. The formalism of the nonequilibrium thermo field dynamics was 
applied to the description of a hydrodynamic state of quantum field systems 
in paper \cite{c26}. Generalized transport equations for nonequilibrium 
quantum systems, specifically for kinetic and hydrodynamic stages, were 
obtained in \cite{c27} on the basis of the thermo field dynamics conception 
\cite{c31,c32} using the nonequilibrium statistical operator method 
\cite{c21,c33,c34}. In this approach, similarly to \cite{c20,c21}, the 
decisive role is that a set of the observed quantities is included in the 
description of the nonequilibrium process. For these quantities one finds 
generalized transport equations which should agree with nonequilibrium 
thermodynamics at controlling the local conservation laws for the 
particles-number density, momentum and energy.  It gives substantial 
advantages over the nonequilibrium Green function technique 
\cite{c14,c15,c16}, which quite well describes excitation spectra, but
practically does not describe nonequilibrium thermodynamics, and has
problems with the local conservation laws control and the generalized 
transport coefficients calculation.

In this paper we consider the kinetics and  hydrodynamics of highly
nonequilibrium and strongly coupled quantum systems using the nonequilibrium 
thermo field dynamics on the basis of the D.N.Zubarev nonequilibrium 
statistical operator method \cite{c27}. Within this method we consider a 
description of the kinetics and hydrodynamics of dense quantum nuclear 
systems with strongly coupled states. The nonequilibrium thermo field 
dynamics on the basis of the nonequilibrium statistical operator method 
constitutes section 2. Thermo field dynamics formalism, superoperators and 
state vectors in the Liouville thermo field space as well as nonequilibrium 
statistical operator and projection operators in thermo field space are 
considered here. A nonequilibrium thermo vacuum state vector is obtained 
here in view of equations for the generalized hydrodynamics of dense 
quantum systems. Transport equations of a consistent description of the 
kinetics and hydrodynamics in thermo field representation are obtained in 
section 3. We mean that these equations are applied to dense quantum 
systems where strong coupled states can appear. This item implies, as one 
of the approaches, to investigate a nonequilibrium nuclear matter 
\cite{c8,c9}.

\section{Nonequilibrium thermo field dynamics on the basis of Zubarev's 
method of nonequilibrium statistical operator}

Let us consider a quantum system of $N$ interacting bosons or fermions. The
Hamiltonian of this system is expressed via creation $\av_l$ and
annihilation $\ai_l$ operators of the corresponding statistics:
%
%	2.1
%
\be
H=H(a^{\dag},a).\label{e2.1}
\ee
Operators $\av_l$, $\ai_l$ satisfy the commutation relations:
%
%	2.2
%
\be
[\ai_l,\av_j]_\sigma=\delta_{lj},\quad
[\ai_l,\ai_j]_\sigma=[\av_l,\av_j]_\sigma=0,\label{e2.2}
\ee
where $[A,B]_\sigma=AB-\sigma BA$, $\sigma=+1$ for bosons and $\sigma=-1$
for fermions.

The nonequilibrium state of such a system is completely described by the 
nonequilibrium statistical operator $\vrho(t)$. This operator satisfies the
quantum Liouville equation
%
%	2.3
%
\be
\ddt\vrho(t)-\frac{1}{\im\hbar}[H,\vrho(t)]=0.\label{e2.3}
\ee
The nonequilibrium statistical operator $\vrho(t)$ allows us to calculate
the average values of operators $A$
%
%	2.4
%
\be
\la A\ra^t=\Sp\lp A\vrho(t)\rp,\label{e2.4}
\ee
which can be observable quantities describing the nonequilibrium state
of the system (for example, a hydrodynamic state is described by the average
values of operators of particle number, momentum and energy densities).

The main idea of thermo field dynamics \cite{c31,c32} and its nonequilibrium
formulation \cite{c35,c36,c37} consists in doing the calculation of average 
values \refp{e2.4} with the help of the so-called nonequilibrium thermo
vacuum state vector:
%
%	2.5
%
\be
\la A\ra^t=\la\la 1|A\vrho(t)\ra\ra=\la\la 1|\hat{A}|\vrho(t)\ra\ra,
\label{e2.5}
\ee
where $\hat{A}$ is a superoperator which acts on state $|\vrho(t)\ra\ra$.
Nonequilibrium thermo vacuum state vector $|\vrho(t)\ra\ra$ satisfies the
Schr\"odinger equation. Starting from equation \refp{e2.3}, we obtain the
relation
\[
\ddt|\vrho(t)\ra\ra-
\left.\left|\frac{1}{\im\hbar}[H,\vrho(t)]\right\ra\!\!\right\ra=0,
\]
or, opening commutator,
%
%	2.6
%
\be
\ddt|\vrho(t)\ra\ra-\frac{1}{\im\hbar}\bar{H}|\vrho(t)\ra\ra=0.\label{e2.6}
\ee
Here the ``total'' Hamiltonian $\bar{H}$ reads:
%
%	2.7
%
\be
\bar{H}=\hat{H}-\tilde{H},\label{e2.7}
\ee
and it is known that $\la\la 1|\bar{H}=0$; $\hat{H}=H(\hav,\hat{a})$,
$\tilde{H}=H^*(\tav,\tilde{a})$ are superoperators which consist of
creation and annihilation superoperators without and with a tilde, and which
represent the thermal Liouville space~\cite{c38,c39}. Superoperators
$\hat{H}$ and $\tilde{H}$ are defined in accordance with the relations:
%
%	2.8
%
\begin{equation}
\begin{array}{l}
|H\vrho(t)\ra\ra=\hat{H}|\vrho(t)\ra\ra,\\
|\vrho(t)H\ra\ra=\tilde{H}|\vrho(t)\ra\ra.\\
\end{array}
\label{e2.8}
\end{equation}
Hence, it appears that at going from the quantum Liouville equation
\refp{e2.3} for non\-equi\-li\-brium statistical operator $\vrho(t)$ to the
Schr\"odinger equation \refp{e2.6} for non\-equi\-lib\-ri\-um thermo vacuum
state vector $|\vrho(t)\ra\ra$, according to \refp{e2.5}, the number of
creation and annihilation operators is doubled. Superoperators
$\hav_l$, $\hai_j$, $\tav_l$, $\tai_j$ satisfy the same commutation
relations as for operators $\av_l$, $\ai_j$ \refp{e2.2}:
%
%	2.9
%
\be
\ba{lcl}
[\hai_l,\hav_j]_\sigma=[\tai_l,\tav_j]_\sigma=\delta_{lj},&\quad&
[\hai_l,\tai_j]_\sigma=[\hav_l,\tav_j]_\sigma=0,\\ [1ex]
[\hai_l,\hai_j]_\sigma=[\hav_l,\hav_j]_\sigma=0,&\quad&
[\tai_l,\tai_j]_\sigma=[\tav_l,\tav_j]_\sigma=0.\\
\ea
\label{e2.9}
\ee
Annihilation superoperators $\hai_l$, $\tai_l$ are defined in accordance with
their action on the vacuum state -- the supervacuum \cite{c38}
%
%	2.10
%
\be
\hai_l|00\ra\ra=\tai_l|00\ra\ra=0,\label{e2.10}
\ee
where $|00\ra\ra=||0\ra\la 0|\ra\ra$ is a supervacuum; and it is 
known that $\hai_l|0\ra=\ai_l|0\ra=0$, and $\la 0|\tai_l=0$, i.e. a
supervacuum $|00\ra\ra$ is an orthogonalized state of two va\-cu\-um states
$\la 0|$ and $|0\ra$. Taking into account commutation relations \refp{e2.9},
\refp{e2.10}, one can introduce unit vectors
$| 1\ra\ra=|\sum_l|l\ra\la l|\ra\ra$ and
$\la\la 1|=\la\la\sum_l|l\ra\la l||$ in the following forms:
%
%	2.11
%
\be
\begin{array}{l}
|1\ra\ra=\exp\lc\sum\nolimits_l\hav_l\tav_l\rc|00\ra\ra,\\
\la\la 1|=\la\la 00|\exp\lc\sum\nolimits_l\tai_l\hai_l\rc.\\
\end{array}
\label{e2.11}
\ee
With the help of these expressions one can find relations between the 
action of superoperators $\hav_l$, $\hai_j$, $\tav_l$, $\tai_j$
%
%	2.12
%
\be
\begin{array}{rcrlcl}
\hai_l|1\ra\ra&=&\tav_l|1\ra\ra,\quad&\la\la 1|\hav_l&=&\la\la 1|\tai_l,\\
\hav_l|1\ra\ra&=&\sigma\tai_l|1\ra\ra,\quad&\la\la 1|\hai_l&=&\la\la
1|\tav_l\sigma.
\end{array}
\label{e2.12}
\ee
In such a way, in the thermal field dynamics formalism \cite{c31,c32} the
number of operators is doubled by introducing both without tilde and
tildian operators $A(\hav,\hat{a})$, $\tilde{A}(\tav,\tilde{a})$ for which 
the following properties take place:
%
%	2.13
%
\be
\begin{array}{rcl}
\widetilde{A_1A_2}&=&\tilde{A}_1\tilde{A}_2,\quad\tilde{\tilde{A}}=A,\\
\widetilde{c_1A_1+c_2A_2}&=&c_1^*\tilde{A}_1+c_2^*\tilde{A}_2,\\
|A\ra\ra&=&\hat{A}|1\ra\ra,\\
|A_1A_2\ra\ra&=&\hat{A}_1|A_2\ra\ra.\\
\end{array}
\label{e2.13}
\ee
Here $^*$ denotes a complex conjugation. Some detailed description of the 
properties of superoperators $\hav_l$, $\hai_j$, $\tav_l$, $\tai_j$, as well
as a thermal Liouville space is given in papers \cite{c31,c32,c38,c39}.

The nonequilibrium thermo vacuum state vector is normalized
%
%	2.14
%
\be
\la\la 1|\vrho(t)\ra\ra=\la\la 1|\hat{\vrho}(t)|1\ra\ra=1,\label{e2.14}
\ee
where $\hat{\vrho}(t)$ is a nonequilibrium statistical superoperator. It
depends on superoperators $\av_l$, $\ai_l$:
$\hat{\vrho}(t)\equiv\vrho\lp\hav,\hat{a};t\rp$, and, it is known that the 
corresponding tildian superoperator
$\tilde{\vrho}(t)\equiv\vrho^{\dag}(\tav,\tilde{a};t)$ depends on 
superoperators $\tav_l$, $\tai_l$.

To solve the Schr\"odinger equation \refp{e2.6} a boundary condition
should be given. Following the nonequilibrium statistical operator method
\cite{c21,c27,c33,c34}, let us find a solution to this equation in a
form, which depends on time via some set of observable quantities only. It
means that this set is sufficient for the description of a nonequilibrium
state of a system and does not depend on the initial moment of time. The
solution to the Schr\"odinger equation, which satisfies the following
boundary condition
%
%	3.1
%
\be
|\vrho(t)\ra\ra_{t=t_0}=|\vrhoq(t_0)\ra\ra,\label{e3.1}
\ee
reads:
%
%	3.2
%
\be
|\vrho(t)\ra\ra=\exp\lc(t-t_0)\frac{1}{\im\hbar}\bar{H}\rc
	|\vrhoq(t_0)\ra\ra.\label{e3.2}
\ee
We will consider times $t\gg t_0$, when the details of the initial state 
become inessential. To avoid the dependence on $t_0$, let us average the 
solution \refp{e3.2} on the initial time moment in the range between $t_0$ 
and $t$ and make the limiting transition $t_0-t\to-\infty$. We will obtain 
\cite{c27}:
%
%	3.3
%
\be
|\vrho(t)\ra\ra=\veps\int_{-\infty}^0\d t'\;\e^{\veps t'}
	\e^{-\frac{1}{\im\hbar}\bar{H}t}|\vrhoq(t+t')\ra\ra,\label{e3.3}
\ee
where $\veps\to+0$ after the thermodynamic limiting transition. Solution 
\refp{e3.3}, as it can be shown by its direct dif\-fe\-ren\-ti\-a\-tion
with respect to time $t$, satisfies the Schr\"odin\-ger equation with a
small source in the right-hand side:
%
%	3.4
%
\be
\lp\ddt-\frac{1}{\im\hbar}\bar{H}\rp|\vrho(t)\ra\ra=
	-\veps\Big(|\vrho(t)\ra\ra-|\vrhoq(t)\ra\ra\Big).\label{e3.4}
\ee
This source selects retarded solutions which correspond to a shortened
description of the nonequilibrium state of a system, $|\vrhoq(t)\ra\ra$ is a
thermo vacuum quasiequilibrium state vector
%
%	3.5
%
\be
|\vrhoq(t)\ra\ra=\hvrhoq(t)|1\ra\ra.\label{e3.5}
\ee
Similarly to \refp{e2.14}, it is normalized by the rule
%
%	3.6
%
\be
\la\la 1|\vrhoq(t)\ra\ra=\la\la 1|\hvrhoq(t)|1\ra\ra=1,\label{e3.6}
\ee
where $\hvrhoq(t)$ is a quasiequilibrium statistical superope\-ra\-tor. The 
quasiequilibrium ther\-mo vacuum state vector of a system is introduced in 
the following way. Let $\la p_n\ra^t=\la\la 1|\hat{p}_n|\vrho(t)\ra\ra$ be 
a set of observable quantities which describe the nonequilibrium state of 
a system. $p_n$ are operators which consist of the creation and annihilation
operators defined in \refp{e2.2}. Quasiequilibrium statistical operator
$\vrhoq(t)$ is defined from the condition of informational entropy $S_{\rm
inf}$ extremum (maximum) at additional conditions of prescribing the average
values $\la p_n\ra^t$ and conservation of normalization condition
\refp{e3.6} \cite{c21,c34}:
%
%	3.7
%
\be
\begin{array}{rcl}
\ds\vrhoq(t)&=&\ds\exp\lc-\Phi(t)-\sum\nolimits_nF_n^*(t)p_n\rc,\\ [1ex]
\ds\Phi(t)&=&\ds\ln\Sp\exp\lc-\sum\nolimits_nF_n^*(t)p_n\rc,\\
\end{array}
\label{e3.7}
\ee
where $\Phi(t)$ is the Massieu-Planck functional. A summation on $n$ can
designate a sum with respect to the wave-vector $\bmv{k}$, the kind of 
particles and a line of other quantum numbers, a spin for example. 
Parameters $F_n(t)$ are defined from the conditions of self-consistency:
%
%	3.8
%
\be
\la p_n\ra^t=\ds\la p_n\ra^t_{\rm q},\qquad
	\la\ldots\ra^t_{\rm q}=\ds\Sp\Big(\ldots\vrhoq(t)\Big).
	\label{e3.8}
\ee
According to \refp{e2.5}, let us write these conditions of self-consistency 
in the following form:
%
%	3.9
%
\be
\la\la 1|\hat{p}_n|\vrho(t)\ra\ra=\la\la 1|\hat{p}_n|\vrhoq(t)\ra\ra.
\label{e3.9}
\ee
Taking into account behaviours \refp{e2.13}, we have:
%
%	3.10
%
\be
|\vrhoq(t)\ra\ra=\hvrhoq(t)|1\ra\ra=\tilde{\vrho}^{\dag}_{\rm q}(t)|1\ra\ra,
\label{e3.10}
\ee
where
%
%	3.11
%
\be
\begin{array}{lcl}
\ds\hvrhoq(t)&=&\ds\exp\lc-\Phi(t)-\sum\nolimits_nF_n^*(t)\hat{p}_n\rc,\\ 
	[1ex]
\ds\tilde{\vrho}_{\rm q}^{\dag}(t)&=&
	\ds\exp\lc-\Phi(t)-\sum\nolimits_nF_n(t)\tilde{p}_n\rc\\
\end{array}
\label{e3.11}
\ee
are quasiequilibrium statistical superoperators which contain 
superoperators $\hat{p}_n$ and $\tilde{p}_n$, correspondingly:
%
%	3.12
%
\be
\begin{array}{lcl}
\hat{p}_n&=&p_n(\hav,\hat{a}),\\
\tilde{p}_n&=&p_n^*(\tav,\tilde{a}).\\
\end{array}
\label{e3.12}
\ee
If self-consistency condition \refp{e3.9} realizes, we shall have the
following relations (at fixed corresponding parameters):
%
%	3.13
%
\be
\begin{array}{l}
\ds\frac{\delta\Phi(t)}{\delta F_n^*(t)}=\la\la 1|\hat{p}_n|\vrhoq(t)\ra\ra
	=\la\la 1|\hat{p}_n|\vrho(t)\ra\ra,\\\\
\ds\frac{\delta\Phi(t)}{\delta F_n(t)}=\la\la 1|\tilde{p}_n|\vrhoq(t)\ra\ra
	=\la\la 1|\tilde{p}_n|\vrho(t)\ra\ra.\\
\end{array}
\label{e3.13}
\ee
Relations \refp{e3.13} show that parameters $F_n^*(t)$, $F_n(t)$ are
conjugated to the averages $\la\la 1|\hat{p}_n|\vrho(t)\ra\ra$ and
$\la\la 1|\tilde{p}_n|\vrho(t)\ra\ra$, correspondingly. On the other hand,
with the help of $|\vrhoq(t)\ra\ra$ and self-consistency conditions
\refp{e3.9} we can define the entropy of the system state:
%
%	3.14
%
\be
S(t)=-\la\la 1|\big(\ln\vrhoq(t)\big)\vrhoq(t)\ra\ra=
\Phi(t)+\sum\nolimits_nF_n^*(t)\la\la 1|\hat{p}_n|\vrho(t)\ra\ra.
\label{e3.14}
\ee
The physical meaning of parameters $F_n^*(t)$  can be obtained now on the
basis of the previous relation:
%
%	3.15
%
\be
F_n^*(t)=\frac{\delta S(t)}{\delta\la\la 1|\hat{p}_n|\vrho(t)}.\label{e3.15}
\ee

Now the auxiliary quasiequilibrium thermo vacuum state vector
$|\vrhoq(t)\ra\ra$ is defined. Let us represent solution \refp{e3.3} of
the Schr\"odinger equation \refp{e2.6} in a form which is more convenient
for the construction of transport equations for averages
$\la\la 1|\hat{p}_n|\vrho(t)\ra\ra$. We shall start from the Schr\"odinger
equation with a small source \refp{e3.4}. Let us rebuild this equation by
introducing $\vt|\vrho(t)\ra\ra=|\vrho(t)\ra\ra-|\vrhoq(t)\ra\ra$:
%
%	4.1
%
\be
\lp\ddt-\frac{1}{\im\hbar}\bar{H}+\veps\rp\vt|\vrho(t)\ra\ra=
-\lp\ddt-\frac{1}{\im\hbar}\bar{H}\rp|\vrhoq(t)\ra\ra.\label{e4.1}
\ee
The calculation of time derivation of $|\vrhoq(t)\ra\ra$ in the right-hand
side of equation \refp{e4.1} is equivalent to the introduction of the
Kawasaki-Gunton projection operator $\SP_{\rm q}(t)$ \cite{c27} in thermo
field representation:
%
%	4.2
%
\be
\ddt|\vrhoq(t)\ra\ra=\SP_{\rm q}(t)\frac{1}{\im\hbar}\bar{H}|\vrho(t)\ra\ra,
\label{e4.2}
\ee
%
%	4.3
%
\bea
\lefteqn{\ds\SP_{\rm q}(t)\big(|\ldots\ra\ra\big)=|\vrhoq(t)\ra\ra+{}}
	\label{e4.3}\\
\lefteqn{\ds\sum_n\frac{\delta|\vrhoq(t)\ra\ra}
	{\delta\la\la 1|\hat{p}_n|\vrho(t)\ra\ra}
	\la\la 1|\hat{p}_n|\ldots\ra\ra-
	\sum_n\frac{\delta|\vrhoq(t)\ra\ra}
	{\delta\la\la 1|\hat{p}_n|\vrho(t)\ra\ra}
	\la\la 1|\hat{p}_n|\ldots\ra\ra\la\la 1|\ldots\ra\ra.}\nonumber
\eea
Projection operator $\SP_{\rm q}(t)$ acts on state vectors $|\ldots\ra\ra$
only and has all the operator properties:
\[
\begin{array}{lcl}
\ds\SP_{\rm q}(t)|\vrho(t')\ra\ra&=&|\vrhoq(t)\ra\ra,\phantom{\intl_0}\\
\ds\SP_{\rm q}(t)|\vrhoq(t')\ra\ra&=&|\vrhoq(t)\ra\ra,\phantom{\intl_0}\\
\ds\SP_{\rm q}(t)\SP_{\rm q}(t')&=&\SP_{\rm q}(t).\phantom{\intl_0}\\
\end{array}
\]
Taking into account condition
$\SP_{\rm q}(t)\frac{1}{\im\hbar}\bar{H}\vt|\vrho(t)\ra\ra=0$, one may
rewrite equation \refp{e4.1}, after simple reductions, in a form:
%
%	4.4
%
\be
\lp\ddt-\Big(1-\SP_{\rm q}(t)\Big)\frac{1}{\im\hbar}\bar{H}+\veps\rp
	\vt|\vrho(t)\ra\ra=
	\ds\Big(1-\SP_{\rm q}(t)\Big)\frac{1}{\im\hbar}\bar{H}
	|\vrhoq(t)\ra\ra.\label{e4.4}
\ee
The formal solution to this equation reads:
\[
\vt|\vrho(t)\ra\ra=\int_{-\infty}^t\d t'\;\e^{\veps(t'-t)}T(t,t')
	\Big(1-\SP_{\rm q}(t')\Big)\frac{1}{\im\hbar}\bar{H}
	|\vrhoq(t')\ra\ra,\qquad\text{or}
\]
%
%
%
%or
%
%	4.5
%
\be
|\vrho(t)\ra\ra=|\vrhoq(t)\ra\ra+\int_{-\infty}^t\d t'\;\e^{\veps(t'-t)}
	T(t,t')\Big(1-\SP_{\rm q}(t')\Big)\frac{1}{\im\hbar}\bar{H}
	|\vrhoq(t')\ra\ra,\label{e4.5}
\ee
where
%
%	4.6
%
\be
T(t,t')=\exp_+\lc\int_{t'}^t\d t'\;
	\Big(1-\SP_{\rm q}(t')\Big)\frac{1}{\im\hbar}\bar{H}\rc
	\label{e4.6}
\ee
is an evolution operator with projection consideration, and $\exp_+$ is an 
ordered exponent. Then, let us consider expression
$\Big(1-\SP_{\rm q}(t')\Big)\frac{1}{\im\hbar}\bar{H}|\vrhoq(t)\ra\ra$ in
the right-hand side of \refp{e4.1}. The action of
$\frac{1}{\im\hbar}\bar{H}$ and $\Big(1-\SP_{\rm q}(t')\Big)$ on
$|\vrhoq(t)\ra\ra$ can be represented in the form:
%
%	4.7
%
\be
\fl\Big(1-\SP_{\rm q}(t')\Big)\frac{1}{\im\hbar}\bar{H}|\vrhoq(t)\ra\ra=
	\sum_nF_n^*(t)\left.\left|
	\mbox{$\intl_0^1$}\d\tau\;\vrhoq^\tau(t)
	\Big(1-\SP(t')\Big)\dot{p}_n\vrhoq^{1-\tau}(t)\Ra\!\!\Ra\!,
	\label{e4.7}
\ee
where $\dot{p}_n$ and $\SP(t)$ read:
%
%	4.8, 4.9
%
\bea
\dot{p}_n&=&-\frac{1}{\im\hbar}[H,p_n],\label{e4.8}\\
\SP(t)p&=&\la\la 1|\hat{p}|\vrhoq(t)\ra\ra+
	\sum_n\frac{\delta\la\la 1|\hat{p}|\vrhoq(t)\ra\ra}
	{\delta\la\la 1|\hat{p}_n|\vrho(t)\ra\ra}
	\Big(p_n-\la\la 1|\hat{p}_n|\vrho(t)\ra\ra\Big).\label{e4.9}
\eea
Here $\SP(t)$ is a generalized Mori projection operator in thermo field
representation. It acts on operators and has the following properties:
%
%	4.10
%
\be
\begin{array}{lcl}
\SP(t)p_n&=&p_n,\\
\SP(t)\SP(t')&=&\SP(t).\\
\end{array}
\label{e4.10}
\ee
Let us substitute now \refp{e4.7} into \refp{e4.5} and, as a result, we will
obtain an expression for the nonequilibrium thermo vacuum state of a system:
%
%	4.11
%
\be
\fl\ds|\vrho(t)\ra\ra=|\vrhoq(t)\ra\ra+
\sum_n\intl_{-\infty}^t\d t'\e^{\veps(t'-t)}T(t,t')
	\left.\left|\mbox{$\intl_0^1$}\d\tau\vrhoq^\tau(t')J_n(t')
	\vrhoq^{1-\tau}(t')\Ra\!\!\Ra\! F_n^*(t').\label{e4.11}
\ee
%
%
%
%Here
%
%	4.12
%
\be
J_n(t)=\Big(1-\SP(t)\Big)\dot{p}_n\label{e4.12}
\ee
are generalized flows.

Let us obtain now transport equations for averages
$\la\la 1|\hat{p}_n|\vrho(t)\ra\ra$ in thermo field representation with the
help of nonequilibrium thermo vacuum state vector $|\vrho(t)\ra\ra$
\refp{e4.11}. To achieve this we will use the equality
%
%	4.13
%
\be
\ddt\la\la 1|\hat{p}_n|\vrho(t)\ra\ra=
	\la\la 1|\dot{\hat{p}}_n|\vrho(t)\ra\ra=
	\la\la 1|\dot{\hat{p}}_n|\vrhoq(t)\ra\ra+
	\la\la J_n(t)|\vrho(t)\ra\ra.\label{e4.13}
\ee
By making use of $|\vrho(t)\ra\ra$ in \refp{e4.11} in averaging the last 
term, we obtain transport equations for $\la\la 1|\hat{p}_n|\vrho(t)\ra\ra$:
%
%	4.14
%
\bea
\lefteqn{\ds\ddt\la\la 1|\hat{p}_n|\vrho(t)\ra\ra=
	\la\la 1|\dot{\hat{p}}_n|\vrhoq(t)\ra\ra+{}}\label{e4.14}\\
\lefteqn{\ds\sum_{n'}\intl_{-\infty}^t\d t'\;\e^{\veps(t'-t)}
	\La\!\!\La J_n(t)T(t,t')\left|\mbox{$\intl_0^1$}\d\tau\;
	\vrhoq^\tau(t')J_{n'}(t')\vrhoq^{1-\tau}(t')\Ra\!\!\Ra\right.
	F_{n'}^*(t'),}\nonumber
\eea
where $\dot{\hat{p}}_n=-\frac{1}{\im\hbar}\bar{H}\hat{p}_n$.
Relations \refp{e4.14} are treated as a general form of transport equations
for average values of a shortened description. These equations can be
applied to completely actual problems.

\section{Transport equations of dense quantum systems with coupled states}

We will consider a quantum field system in which coupled states can
appear between the particles. Let us introduce annihilation and creation
operators of a coupled state $(A\alpha)$ with $A$-particle:
%
%	8.1
%
\be
\ba{l}
\ds\ai_{A\alpha}(\bmv p)=\sum_{1,\ldots,A}\mPsi_{A\alpha\bmv p}
	(1,\ldots,A)a(1)\ldots a(A),\\
\ds\av_{A\alpha}(\bmv p)=\sum_{1,\ldots,A}\mPsi^*_{A\alpha\bmv p}
	(1,\ldots,A)\av(1)\ldots\av(A),\\
\ea\label{e8.1}
\ee
where $\mPsi_{A\alpha\bmv p}(1,\ldots,A)$ is a self-function of the 
$A$-particle coupled state, $\alpha$ denotes internal quantum numbers 
(spin, etc.), $\bmv{p}$ is a particle momentum, the sum covers the 
particles. Annihilation and creation operators $a(j)$ and $\av(j)$ satisfy 
the following commutation relations:
%
%	8.2
%
\be
[a(l),\av(j)]_{\sigma}=\delta_{l,j},\qquad
	[a(l),a(j)]_{\sigma}=[\av(l),\av(j)]_{\sigma}=0,
	\label{e8.2}
\ee
where $\sigma$-commutator is determined by $[a,b]_\sigma=ab-\sigma ba$ with
$\sigma=\pm 1$: $+1$ for bosons and $-1$ for fermions.

The Hamiltonian of such a system can be written in the form:
%
%	8.3
%
\bea
\fl\lefteqn{\ds H=\sum_{A,\alpha}\int\frac{\d\bmv{p}\d\bmv{q}}
	{(2\pi\hbar)^6}\frac{p^2}{2m_A}
	\av_{A\alpha}\lp\bmv{p}-\frac{\bmv{q}}{2}\rp
	\ai_{A\alpha}\lp\bmv{p}+\frac{\bmv{q}}{2}\rp+{}}
	\label{e8.3}\\
\fl\lefteqn{\ds\frac 12\sum_{A,B}\sum_{\alpha,\beta}
	\int\frac{\d\bmv{p}\d\bmv{p}'\d\bmv{q}}{(2\pi\hbar)^9}V_{AB}(\bmv{q})
	\av_{A\alpha}\lp\bmv{p}+\frac{\bmv{q}-\bmv{p}'}{2}\rp
	\hat{n}_{B\beta}^{\php}(\bmv{q})
	\ai_{A\alpha}\lp\bmv{p}-\frac{\bmv{q}-\bmv{p}'}{2}\rp,}
	\nonumber
\eea
where $V_{AB}(\bmv{q})$ is interaction energy between $A$- and
$B$-particle coupled states, $\bmv{q}$ is a wavevector. Annihilation and 
creation operators $\ai_{A\alpha}(\bmv{p})$ and $\av_{A\alpha}(\bmv{p})$ 
satisfy the following commutation relations:
%
%	8.4
%
\be
\ba{l}
\ds[\ai_{A\alpha}(\bmv{p}),\av_{B\beta}(\bmv{p}')]_{\sigma}=
	\delta_{A,B}\delta_{\alpha,\beta}\delta(\bmv{p}-\bmv{p}'),\\
\ds[\ai_{A\alpha}(\bmv{p}),\ai_{B\beta}(\bmv{p}')]_{\sigma}=
	[\av_{A\alpha}(\bmv{p}),\av_{B\beta}(\bmv{p}')]_{\sigma}=0.\\
\ea\label{e8.4}
\ee
$\hat{n}_{B\beta}(\bmv{q})$ in (\ref{e8.3}) is a Fourier transform of
the $B$-particle density operator:
\[
\hat{n}_{B\beta}(\bmv{q})=\int\frac{\d\bmv{p}}{(2\pi\hbar)^3}\,
	\av_{\bmv{p}-\frac{\bmv{q}}{2}}\ai_{\bmv{p}+\frac{\bmv{q}}{2}}.
\]

As parameters $\la\la 1|\hat{p}_n|\vrho(t)\ra\ra$ of a shortened 
description for the consistent description of the kinetics and 
hydrodynamics of a system, where coupled states between the particles can 
appear, let us choose nonequilibrium distribution functions of $A$-particle 
coupled states in thermo field representation
%
%	8.5
%
\be
\la\la 1|\hat{n}_{A\alpha}^{\php}(\bmv{r},\bmv{p})|\vrho(t)\ra\ra=
	f_{A\alpha}^{\php}(\bmv{r},\bmv{p};t)=f_{A\alpha}^{\php}(x;t),\quad
	x=\{\bmv{r},\bmv{p}\},\label{e8.5}
\ee
here $f_{A\alpha}^{\php}(x;t)$ is a Wigner function of the $A$-particle 
coupled state where
%
%	8.6
%
\be
\hat{n}_{A\alpha}^{\php}(\bmv{r},\bmv{p})\equiv\hat{n}_{A\alpha}^{\php}(x)=
	\int\frac{\d\bmv{q}}{(2\pi\hbar)^3}\,
	\e^{-\frac{1}{\im\hbar}\bmv{q}\cdot\bmv{r}}
	\hav_{A\alpha}\lp\bmv{p}-\frac{\bmv{q}}{2}\rp
	\hai_{A\alpha}\lp\bmv{p}+\frac{\bmv{q}}{2}\rp
	\label{e8.6}
\ee
is the Klimontovich density operator; and the average value of the total 
energy density operator
%
%	8.7
%
\be
\la\la 1|\hat{H}(\bmv{r})|\vrho(t)\ra\ra=\la\la 1|H(\bmv{r})\vrho(t)\ra\ra.
	\label{e8.7}
\ee
By this $\int\d\bmv{r}\;H(\bmv{r})=H$, $\hat{H}(\bmv{r})$ is a 
superoperator of the total energy density which is constructed on 
annihilation and creation superoperators $\hai{A\alpha}(\bmv{p})$ and 
$\hav_{A\alpha}(\bmv{p})$. The latter satisfy commutation relations
(\ref{e8.4}). Following \cite{c27} and (\ref{e3.7}), one can rewrite 
quasiequilibrium statistical operator $\hat{\vrho}_{\rm q}(t)$, 
$|\vrho_{\rm q}(t)\ra\ra=\hat{\vrho}_{\rm q}(t)|1\ra\ra$ for the mentioned
parameters of a shortened description in the form:
%
%	8.8
%
\bea
\fl\lefteqn{\ds\hat{\vrho}_{\rm q}(t)=\exp\lc-\Phi^*(t)-
	\int\d\bmv{r}\;\beta(\bmv{r};t)\lp\hat{H}(\bmv{r})-\sum_{A,\alpha}
	\int\frac{\d\bmv{p}}{(2\pi\hbar)^3}\,
	\mu_{A\alpha}^{\php}(x;t)\hat{n}_{A\alpha}^{\php}(x)\rp\rc,}
	\nonumber\\
	\label{e8.8}
\eea
where Lagrange multipliers $\beta(\bmv{r};t)$ and
$\mu_{A\alpha}^{\php}(x;t)$ can be found from the self-con\-sis\-ten\-cy
conditions (\ref{e3.8}), correspondingly:
%
%	8.9, 8.10
%
\bea
\la\la 1|\hat{H}(\bmv{r})|\vrho(t)\ra\ra&=&
	\la\la 1|\hat{H}(\bmv{r})|\vrho_{\rm q}(t)\ra\ra,
	\label{e8.9}\\
\la\la 1|\hat{n}_{A\alpha}^{\php}(x)|\vrho(t)\ra\ra&=&
	\la\la 1|\hat{n}_{A\alpha}^{\php}(x)|\vrho_{\rm q}(t)\ra\ra,
	\label{e8.10}
\eea
$\Phi^*(t)$ is the Massieu-Planck functional and it can be defined from the
normalization condition \refp{e3.6}:
%
%	8.11
%
\bea
\fl\lefteqn{\ds\Phi^*(t)=\ln\La\!\!\La 1\left|
	\exp\lc-\!\!\int\d\bmv{r}\;\beta(\bmv{r};t)
	\lp\hat{H}(\bmv{r})-\sum_{A,\alpha}
	\int\frac{\d\bmv{p}}{(2\pi\hbar)^3}
	\mu_{A\alpha}^{\php}(x;t)\hat{n}_{A\alpha}^{\php}(x)
	\rp\rc\right.\Ra\!\!\Ra.}
	\nonumber\\
	\label{e8.11}
\eea
Using now the general structure of nonequilibrium thermo field dynamics
\refp{e4.1}--\refp{e4.14}, one can obtain a set of generalized transport
equations for $A$-particle Wigner distribution functions and the average
interaction energy:
%
%	8.12
%
\bea
\fl&&\frac{\partial}{\partial t}\la\la 1|\hat{n}_{A\alpha}^{\php}(x)
	|\vrho(t)\ra\ra=\la\la 1|\dot{\hat{n}}_{A\alpha}^{\php}(x)
	|\vrho_{\rm q}(t)\ra\ra+
	\int\d\bmv{r}'\int_{-\infty}^t\d t'\;\e^{\veps(t'-t)}
	\vphi^{A\alpha}_{nH}(x,\bmv{r}';t,t')\beta(\bmv{r}';t')
	\nonumber\\
\fl&&+\sum_{B,\beta}\int\d x'\int_{-\infty}^t\d t'\;\e^{\veps(t'-t)}
	\vphi^{A\alpha B\beta}_{nn}(x,x';t,t')
	\beta(\bmv{r}';t')\mu_{B\beta}^{\php}(x';t'),
	\label{e8.12}
\eea
%
%
%
%
%	8.13
%
\bea
\fl&&\frac{\partial}{\partial t}\la\la 1|\hat{H}(\bmv{r})
	|\vrho(t)\ra\ra=\la\la 1|\dot{\hat{H}}(\bmv{r})
	|\vrho_{\rm q}(t)\ra\ra+
	\int\d\bmv{r}'\int_{-\infty}^t\d t'\;\e^{\veps(t'-t)}
	\vphi_{HH}^{\phantom B}(\bmv{r},\bmv{r}';t,t')\beta(\bmv{r}';t')
	\nonumber\\
\fl&&+\sum_{B,\beta}\int\d x'\int_{-\infty}^t\d t'\;\e^{\veps(t'-t)}
	\vphi^{B\beta}_{Hn}(\bmv{r},x';t,t')
	\beta(\bmv{r}';t')\mu_{B\beta}^{\php}(x';t'),
	\label{e8.13}
\eea
where $x'=\{\bmv{r}',\bmv{p}'\}$,
$dx'=(2\pi\hbar)^{-3}d\bmv{r}'\,d\bmv{p}'$. Here
%
%	8.14, 8.15, 8.16, 8.17
%
\bea
\fl\vphi_{nn}^{A\alpha\atop B\beta}(x,x';t,t')&=&
	\La\!\!\La 1\bigg|\hat{J}_{n_{A\alpha}^{\php}}(x,t)T(t,t')
	\bigg|\mbox{$\intl_0^1$}\d\tau\;\vrho_{\rm q}^\tau(t')
	J_{n_{B\beta}^{\php}}(x';t')\vrho_{\rm q}^{1-\tau}(t')\!\Ra\!\!\Ra\!,
	\label{e8.14}\\
\fl\vphi_{nH}^{A\alpha}(x,\bmv{r}';t,t')&=&
	\La\!\!\La 1\bigg|\hat{J}_{n_{A\alpha}^{\php}}(x,t)T(t,t')
	\bigg|\mbox{$\intl_0^1$}\d\tau\;\vrho_{\rm q}^\tau(t')
	J_{H}(\bmv{r}';t')\vrho_{\rm q}^{1-\tau}(t')\Ra\!\!\Ra,
	\label{e8.15}\\
\fl\vphi_{Hn}^{B\beta}(\bmv{r}',x';t,t')&=&
	\La\!\!\La 1\bigg|\hat{J}_{H}(\bmv{r},t)T(t,t')
	\bigg|\mbox{$\intl_0^1$}\d\tau\;\vrho_{\rm q}^\tau(t')
	J_{n_{B\beta}^{\php}}(x';t')\vrho_{\rm q}^{1-\tau}(t')\Ra\!\!\Ra,
	\label{e8.16}\\
\fl\vphi_{HH}^{\phantom B}(\bmv{r},\bmv{r}';t,t')&=&
	\La\!\!\La 1\bigg|\hat{J}_{H}(\bmv{r},t)T(t,t')
	\bigg|\mbox{$\intl_0^1$}\d\tau\;\vrho_{\rm q}^\tau(t')
	J_{H}(\bmv{r}';t')\vrho_{\rm q}^{1-\tau}(t')\Ra\!\!\Ra,
	\label{e8.17}%\label{e2.30}
\eea
are generalized transport cores which describe dissipative processes. In 
these formulae $J$ are generalized flows:
%
%	8.18
%
{\setlength{\arraycolsep}{0.5mm}
\be
\begin{array}{@{}lll}
\ds J_H(\bmv{r};t)&=&\ds\Big(1-{\SP}(t')\Big)\dot{H}(\bmv{r}),\\[0.75ex]
\ds J_{n_{A\alpha}^{\php}}(\bmv{r},\bmv{p};t)&=&\ds\Big(1-{\SP}(t')\Big)
	\dot{n}_{A\alpha}^{\php}(x),\\
\end{array}
\quad
\begin{array}{@{}lll}
\ds\dot{H}(\bmv{r})&=&\ds-\frac{1}{\im\hbar}[H,H(\bmv{r})],\\[1.5ex]
\ds\dot{n}_{A\alpha}^{\php}(\bmv{r},\bmv{p})&=&\ds
	-\frac{1}{\im\hbar}[H,n_{A\alpha}^{\php}(x)],\\
\end{array}
\label{e8.18}
\ee}
\noindent ${\SP}(t)$ is a generalized Mori projection operator in thermo 
field representation. It acts on operators
%
%	8.19
%
\bea
&&{\SP}(t)P=\la\la|\hat{P}|\vrho_{\rm q}(t)\ra\ra+
	\int\d\bmv{r}\;\frac{\delta\la\la 1|\hat{P}|\vrho_{\rm q}(t)\ra\ra}
	{\delta\la\la 1|\hat{H}(\bmv{r})|\vrho(t)\ra\ra}
	\Big(H(\bmv{r})-\la\la 1|\hat{H}(\bmv{r})|\vrho(t)\ra\ra\Big)
	\nonumber\\
&&+\sum_{A,\alpha}\int\frac{\d\bmv{r}\,\d\bmv{p}}{(2\pi\hbar)^3}
	\frac{\delta\la\la 1|\hat{P}|\vrho_{\rm q}(t)\ra\ra}
	{\delta\la\la 1|\hat{n}_{A\alpha}^{\php}(x)|\vrho(t)\ra\ra}
	\Big(n_{A\alpha}^{\php}(x)-
	\la\la 1|\hat{n}_{A\alpha}^{\php}(x)|\vrho(t)\ra\ra\Big)
	\label{e8.19}
\eea
and has all the properties of a projection operator:
\[
\begin{array}{@{}lll@{\qquad}lll}
\ds{\SP}(t)H(\bmv{r})&=&H(\bmv{r}),&
	\ds{\SP}(t){\SP}(t')&=&{\SP}(t),\\[0.75ex]
\ds{\SP}(t)n_{A\alpha}^{\php}(\bmv{r},\bmv{p})&=&
	\ds n_{A\alpha}^{\php}(\bmv{r},\bmv{p}),&
	\ds\Big(1-{\SP}(t)\Big){\SP}(t)&=&0.
\end{array}
\]

The obtained transport equations have the general meaning and can describe 
both weakly and strongly nonequilibrium processes of a quantum system with 
taking into consideration coupled states. In a low density quantum field
Bose- or Fermi-system the influence of the average value of interaction 
energy is substantially smaller than the average kinetic energy, and 
coupled states between the particles are absent. In such a case the set of 
transport equations \refp{e8.12}, \refp{e8.13} is simplified. It transforms 
into a kinetic equation \cite{c27} in thermo field representation for the 
average value of the Klimontovich operator 
$\la\la 1|\hat{n}(x)|\vrho(t)\ra\ra$:
\bean
\fl\lefteqn{\ds\ddt\la\la 1|\hat{n}_{\bmv{k}}(\bmv{p})|\vrho(t)\ra\ra=
	\la\la 1|\dot{\hat{n}}_{\bmv{k}}(\bmv{p})|
	\vrhoq(t)\ra\ra+{}}\\
\fl\lefteqn{\ds\sum_{\gb}\int\d\bmv{p}'
	\int_{-\infty}^t\d t'\;\e^{\veps(t'-t)}
	\La\!\!\La J_n(\bmv{k};t)\left|T(t,t')\left|
	\mbox{$\intl_0^1$}\d\tau\;
	\vrhoq^\tau(t')J_n(\gb;t')
	\vrhoq^{1-\tau}(t')\Ra\!\!\Ra\right.\right.
	b_{-\gb}(\bmv{p}';t').}\nonumber
\eean
Using the projection operators method, this equation was obtained in 
\cite{c24}.

In the next step we will construct such annihilation and creation 
superoperators, for which the quasiequilibrium thermo vacuum state vector 
is a vacuum state. Analysing the structure of quasiequilibrium statistical
superoperator \refp{e8.8}, one can mark out some part which would
correspond to the system of noninteracting quantum $A$-particles. Let us
write $\hat{\vrho}_{\rm q}(t)$ in an evident form and separate terms which
are connected with the interaction energy between the particles:
%
%	8.20
%
\bea
\lefteqn{\ds\hat{\vrho}_{\rm q}(t)=\exp\lc-\Phi^*(t)-
	\int\d\bmv{r}\;\beta(\bmv{r};t)
	\times{}\right.}\label{e8.20}\\
\lefteqn{\ds\left.\sum_{A,\alpha}\int\frac{\d\bmv{r}\d\bmv{p}}{(2\pi\hbar)^3}
	\ls\frac{\bmv{p}^2}{2m_A}\hat{n}_{A\alpha}^{\php}(x)-
	\mu_{A\alpha}^{\php}(x;t)\hat{n}_{A\alpha}^{\php}(x)\rs-
	\int\d\bmv{r}\beta(\bmv{r};t)\hat{H}_{\rm int}(\bmv{r})\right\}.}
	\nonumber
\eea
Using operator equality ($A$ and $B$ are some operators)
\[
\ds\e^{A+B}=\ls1+\int_0^1\d\tau\;\e^{\tau(A+B)}\,B\,\e^{-\tau(A+B)}\rs\e^{A},
\]
the relation for $\hat{\vrho}_{\rm q}(t)$ can be rewritten in the following
form:
%
%	8.21
%
\be
\hat{\vrho}_{\rm q}(t)=\ls 1-\int\d\bmv{r}\;\beta(\bmv{r};t)
	\int_0^1\d\tau\;\hat{\vrho}_{\rm q}^\tau(t)\hat{H}_{\rm int}(\bmv{r})
	\big(\hat{\vrho}_{\rm q}(t)\big)^{-\tau}\rs
	\hat{\vrho}_{\rm q}^0(t),\label{e8.21}
\ee
where
%
%	8.22
%
\be
\hat{\vrho}_{\rm q}^0(t)=\exp\lc\Phi(t)-
	\int\d\bmv{r}\;\beta(\bmv{r};t)
	\sum_{A,\alpha}\int\frac{\d\bmv{p}}{(2\pi\hbar)^3}
	b_{A\alpha}^{\php}(x;t)\hat{n}_{A\alpha}^{\php}(x)\rc,
	\label{e8.22}
\ee
%
%
%
%
%	8.23
%
\be
\ds b_{A\alpha}^{\php}(x;t)=
\ls\frac{\bmv{p}^2}{2m_A}\hat{n}_{A\alpha}^{\php}(x)-
\mu_{A\alpha}^{\php}(x;t)\hat{n}_{A\alpha}^{\php}(x)\rs.
\label{e8.23}
\ee
Quasiequilibrium statistical su\-per\-o\-pe\-ra\-tor 
$\hat{\vrho}_{\rm q}^0(t)$
is bilinear on annihilation and creation superoperators
$\hai_{A\alpha}(\bmv{P})$ and $\hav_{A\alpha}(\bmv{P})$, as well as on the
non-perturbed part of Hamiltonian $\bar{H}_0$. One can write the total
quasiequilibrium superoperator as some non-perturbed part of
$\hat{\vrho}_{\rm q}^0(t)$ and the part which describes interaction of
quantum particles in the quasiequilibrium state. Further, we introduce the
following designation:
%
%	8.24
%
\be
\hat{\vrho}_{\rm q}(t)=\hat{\vrho}_{\rm q}^0(t)+\hat{\vrho}_{\rm q}^\prime(t),
\label{e8.24}
\ee
where
%
%	8.25
%
\be
\hat{\vrho}_{\rm q}^\prime(t)=-\int\d\bmv{r}\;\beta(\bmv{r};t)
	\int_0^1\d\tau\;\hat{\vrho}_{\rm q}^\tau(t)\hat{H}_{\rm int}(\bmv{r})
	\big(\hat{\vrho}_{\rm q}(t)\big)^{-\tau}\hat{\vrho}_{\rm q}^0(t).
	\label{e8.25}
\ee
Quasiequilibrium thermo vacuum states $|\hat{\vrho}_{\rm q}(t)\ra\ra$ and
$|\hat{\vrho}_{\rm q}^0(t)\ra\ra$ are not vacuum states for annihilation and
creation superoperators $\hai_{A\alpha}(\bmv{P})$,
$\hav_{A\alpha}(\bmv{P})$, $\tai_{A\alpha}(\bmv{P})$,
$\tav_{A\alpha}(\bmv{P})$. But for $|\hat{\vrho}_{\rm q}^0(t)\ra\ra$ one can
construct new superoperators $\hgi_{A\alpha}(\bmv{P})$,
$\hgv_{A\alpha}(\bmv{P})$, $\tgi_{A\alpha}(\bmv{P})$,
$\tgv_{A\alpha}(\bmv{P})$ as a linear combination of  superoperators
$\hai_{A\alpha}(\bmv{P})$, $\hav_{A\alpha}(\bmv{P})$ and
$\tai_{A\alpha}(\bmv{P})$, $\tav_{A\alpha}(\bmv{P})$ in order to satisfy 
the conditions:
%
%	8.26
%
\be
\ba{lll}
\ds\hgi_{A\alpha}(\bmv{P};t)|\vrho_{\rm q}^0(t)\ra\ra=0,
	&\qquad&\ds\la\la 1|\hgv_{A\alpha}(\bmv{P};t)=0,\\[0.5ex]
\ds\tgi_{A\alpha}(\bmv{P};t)|\vrho_{\rm q}^0(t)\ra\ra=0,
	&\qquad&\ds\la\la 1|\tgv_{A\alpha}(\bmv{P};t)=0.\\
\ea\label{e8.26}
\ee
To achieve this let us consider an action of annihilation superoperators
$\hai_{A\alpha}(\bmv{P};t)$, $\tai_{A\alpha}(\bmv{P};t)$ on quasiequilibrium
state $|\vrho_{\rm q}^0(t_0)\ra\ra$:
%
%	8.27
%
\bea
\ds\hai_{A\alpha}(\bmv{P};t)|\vrho_{\rm q}^0(t_0)\ra\ra&=&
	\phantom{\sigma}f_{A\alpha}(\bmv{P};t-t_0)
	\tav_{A\alpha}(\bmv{P};t)|\vrho_{\rm q}^0(t_0)\ra\ra,
	\nonumber\\
\ds\tai_{A\alpha}(\bmv{P};t)|\vrho_{\rm q}^0(t_0)\ra\ra&=&
	\sigma f_{A\alpha}(\bmv{P};t-t_0)
	\hav_{A\alpha}(\bmv{P};t)|\vrho_{\rm q}^0(t_0)\ra\ra,
	\label{e8.27}%\\
%	\nonumber\\
%\ds\la\la 1|\hai_{A\alpha}(\bmv{P};t)&=&
%	\la\la 1|\tav_{A\alpha}(\bmv{P};t),\nonumber
\eea
where superoperators $\hai_{A\alpha}(\bmv{p};t)$, 
$\hav_{A\alpha}(\bmv{p};t)$, $\tai_{A\alpha}(\bmv{p};t)$, 
$\tav_{A\alpha}(\bmv{p};t)$ are in the Heisenberg representation
\bean
\lefteqn{\ds\hai_{A\alpha}(\bmv{P};t)=
	\e^{-\frac{1}{\im\hbar}\bar{H}_0t}
	\,\hai_{A\alpha}(\bmv{P})\,
	\e^{\frac{1}{\im\hbar}\bar{H}_0t},\qquad
	\tai_{A\alpha}(\bmv{P};t)=
	\e^{-\frac{1}{\im\hbar}\bar{H}_0t}
	\,\tai_{A\alpha}(\bmv{P})\,
	\e^{\frac{1}{\im\hbar}\bar{H}_0t},}\\
\lefteqn{\ds\hav_{A\alpha}(\bmv{P};t)=
	\e^{-\frac{1}{\im\hbar}\bar{H}_0t}
	\,\hav_{A\alpha}(\bmv{P})\,
	\e^{\frac{1}{\im\hbar}\bar{H}_0t},\qquad
	\tav_{A\alpha}(\bmv{P};t)=
	\e^{-\frac{1}{\im\hbar}\bar{H}_0t}
	\,\tav_{A\alpha}(\bmv{P})\,
	\e^{\frac{1}{\im\hbar}\bar{H}_0t},}
\eean
and satisfy commutation relations:
\[
\ba{l}
\ds\ls\hai_{A\alpha}(\bmv{P};t),\hav_{B\beta}(\bmv{P}';t)\rs_\sigma=
	\delta_{A,B}\delta_{\alpha,\beta}\delta(\bmv{P}-\bmv{P}'),\\
\ds\ls\tai_{A\alpha}(\bmv{P};t),\tav_{B\beta}(\bmv{P}';t)\rs_\sigma=
	\delta_{A,B}\delta_{\alpha,\beta}\delta(\bmv{P}-\bmv{P}'),\\
\ds\ls\hai_{A\alpha}(\bmv{P};t),\tai_{B\beta}(\bmv{P}';t)\rs_\sigma=
	\ls\hav_{A\alpha}(\bmv{P};t),\tav_{B\beta}(\bmv{P}';t)\rs_\sigma=0.\\
\ea
\]
It is necessary to note that superoperators $\hat{H}(\bmv{r})$,
$\hat{n}_{A\alpha}^{\php}(x)$ are built on superoperators
$\hai_{A\alpha}(\bmv{p}+\frac{\bmv{q}}{2})$,
$\hav_{A\alpha}(\bmv{p}-\frac{\bmv{q}}{2})$,
$\tai_{A\alpha}(\bmv{p}+\frac{\bmv{q}}{2})$,
$\tav_{A\alpha}(\bmv{p}-\frac{\bmv{q}}{2})$.
Therefore, for convenience here a unit denotion was introduced for arguments
like $\bmv{P}=\bmv{p}\pm\frac{\bmv{q}}{2}$. This should be taken into
account in further calculations where obvious expressions are needed.

%According to general relations of section 6 \refp{e6.7}--\refp{e6.19},
We can introduce new operators
$\hgi_{A\alpha}(\bmv{P};t)$, $\hgv_{A\alpha}(\bmv{P};t)$,
$\tgi_{A\alpha}(\bmv{P};t)$, $\tgv_{A\alpha}(\bmv{P};t)$ via
superoperators
$\hai_{A\alpha}(\bmv{P};t)$, $\hav_{A\alpha}(\bmv{P};t)$,
$\tai_{A\alpha}(\bmv{P};t)$, $\tav_{A\alpha}(\bmv{P};t)$ in the following 
way:
%
%	8.28
%
\bea
\fl\ds\hgi_{A\alpha}(\bmv{P};t)&=&
	\sqrt{1+\sigma n_{A\alpha}^{\php}(\bmv{P};t,t_0)}
	\ls\hai_{A\alpha}(\bmv{P};t)-
	{\frac{n_{A\alpha}^{\php}(\bmv{P};t,t_0)}
	{1+\sigma n_{A\alpha}^{\php}(\bmv{P};t,t_0)}}
	\tav_{A\alpha}(\bmv{P};t)\rs,\nonumber\\
\fl\ds\tgv_{A\alpha}(\bmv{P};t)&=&
	\sqrt{1+\sigma n_{A\alpha}^{\php}(\bmv{P};t,t_0)}
	\ls\tav_{A\alpha}(\bmv{P};t)-
	\sigma\hai_{A\alpha}(\bmv{P};t)\rs.
	\label{e8.28}
\eea
Relations \refp{e8.28} satisfy conditions (8.26). Here
\bean
n_{A\alpha}^{\php}(\bmv{p},\bmv{q};t,t_0)=
	n_{A\alpha}^{\php}(\bmv{P};t,t_0),=\nonumber
\lefteqn{\ds\la\la 1|\tav_{A\alpha}(\bmv{P};t)\tai_{A\alpha}(\bmv{P};t)
	|\vrho_{\rm q}^0(t_0)\ra\ra={}}\\
\lefteqn{\ds\la\la 1|\tav_{A\alpha}(\bmv{p}-\frac{\bmv{q}}{2};t)
	\tai_{A\alpha}(\bmv{p}+\frac{\bmv{q}}{2};t)|\vrho_{\rm q}^0(t_0)
	\ra\ra,}
\eean
is a quasiequilibrium distribution function of $A$-particle coupled
states in momentum space $\bmv{p}$, $\bmv{q}$, which is calculated with the
help of quasiequilibrium thermo vacuum state vector
$\vrho_{\rm q}^0(t_0)\ra\ra$ \refp{e8.22}. Function
$f_{A\alpha}(\bmv{P};t-t_0)$ in formulae \refp{e8.27} is connected
with $n_{A\alpha}(\bmv{P};t,t_0)$ by the relation
\[
f_{A\alpha}(\bmv{P};t-t_0)=\frac{n_{A\alpha}(\bmv{P};t,t_0)}
	{1+\sigma n_{A\alpha}(\bmv{P};t,t_0)}.
\]
Superoperators
$\hgi_{A\alpha}(\bmv{P};t)$ and $\tgi_{A\alpha}(\bmv{P};t)$,
$\hgv_{A\alpha}(\bmv{P};t)$ and $\tgv_{A\alpha}(\bmv{P};t)$ satisfy the
``ca\-no\-ni\-cal'' commutation relations:
%
%	8.29
%
\be
\ba{l}
\ds\ls\hgi_{A\alpha}(\bmv{P};t),\hgv_{B\beta}(\bmv{P}';t)\rs_\sigma=
	\delta_{A,B}\delta_{\alpha,\beta}\delta(\bmv{P}-\bmv{P}'),\\
\ds\ls\tgi_{A\alpha}(\bmv{P};t),\tgv_{B\beta}(\bmv{P}';t)\rs_\sigma=
	\delta_{A,B}\delta_{\alpha,\beta}\delta(\bmv{P}-\bmv{P}'),\\
\ds\ls\hgi_{A\alpha}(\bmv{P};t),\tgi_{B\beta}(\bmv{P}';t)\rs_\sigma=
	\ls\hgv_{A\alpha}(\bmv{P};t),\tgv_{B\beta}(\bmv{P}';t)\rs_\sigma=0.\\
\ea\label{e8.29}
\ee
Inversed transformations to superoperators $\hai_{A\alpha}(\bmv{P};t)$,
$\tav_{A\alpha}(\bmv{P};t)$ are easily obtained from \refp{e8.28}:
%
%	8.30
%
\bea
\fl\ds\hai_{A\alpha}(\bmv{P};t)&=&
	\sqrt{1+\sigma n_{A\alpha}^{\php}(\bmv{P};t,t_0)}
	\ls\hgi_{A\alpha}(\bmv{P};t)+
	{\frac{n_{A\alpha}^{\php}(\bmv{P};t,t_0)}
	{1+\sigma n_{A\alpha}^{\php}(\bmv{P};t,t_0)}}
	\tgv_{A\alpha}(\bmv{P};t)\rs,\nonumber\\
\fl\ds\tav_{A\alpha}(\bmv{P};t)&=&
	\sqrt{1+\sigma n_{A\alpha}^{\php}(\bmv{P};t,t_0)}
	\ls\tgv_{A\alpha}(\bmv{P};t)+
	\sigma\hgi_{A\alpha}(\bmv{P};t)\rs.
	\label{e8.30}
\eea
$\hgi_{A\alpha}(\bmv{P};t)$, $\hgv_{A\alpha}(\bmv{P};t)$,
$\tgi_{A\alpha}(\bmv{P};t)$, $\tgv_{A\alpha}(\bmv{P};t)$ could be
de\-fi\-ned as some operators of annihilation and creation of
$A$-quasiparticle coupled states, for which quasiequilibrium thermo vacuum
state $|\vrho_{\rm q}^0(t_0)\ra\ra$ \refp{e8.22} is a vacuum state. In such
a way, we obtained relations of dynamical reflection of
superoperators $\hai_{A\alpha}(\bmv{P};t)$, $\hav_{A\alpha}(\bmv{P};t)$,
$\tai_{A\alpha}(\bmv{P};t)$, $\tav_{A\alpha}(\bmv{P};t)$ to new
superoperators of ``quasiparticles'' $\hgi_{A\alpha}(\bmv{P};t)$,
$\hgv_{A\alpha}(\bmv{P};t)$, $\tgi_{A\alpha}(\bmv{P};t)$,
$\tgv_{A\alpha}(\bmv{P};t)$.

\section{Conclusions}

A set of transport equations \refp{e8.12}, \refp{e8.13} together with
dynamical reflections \refp{e8.28}, \refp{e8.30} of superoperators in the 
thermo field space constitute the basis for a consistent description of the 
kinetics and hydrodynamics of a dense quantum system with strongly coupled
states. Both strongly and weakly nonequilibrium processes of a nuclear 
matter can be investigated using this approach, in which the particle
interaction is characterized by strongly coupled states, taking into 
account theirs nuclear nature \cite{c1,c2,c8,c9}.

Another problem in the description of quantum kinetic processes of nuclear 
col\-li\-si\-ons should be noted. It is connected with the construction of 
quantum kinetic equations for small times with taking account initial 
states and non-Markovian memory effects. One approach to obtain such 
kinetic equations is developed on the basis of mixed Green functions in 
recent papers by Morozov, R\"opke and others \cite{x,y}. In our approach 
the problem of initial states is connected with the quasiequilibrium thermo 
vacuum state $|\vrhoq(t_0)\ra\ra$ at the initial time $t_0$. Non-Markovian 
memory effects are described here by the generalised memory functions 
$\vphi_{nn}^{A\alpha B\beta}$, $\vphi_{nH}^{A\alpha}$, 
$\vphi_{Hn}^{B\beta}$, $\vphi_{HH}$ (\ref{e8.14})--(\ref{e8.17}), 
correspondingly. These functions are calculated with the help of 
$|\vrho(t)\ra\ra_{t=t_0}$ or $|\vrhoq(t_0)\ra\ra$ (see \cite{z}) and take 
into account both oneparticle and manyparticle processes of energy transfer 
in a system.

In the next paper we will consider in detail weakly nonequilibrium case and 
ob\-tain generalised closed transfer equations for Wigner function 
$f_{A\alpha}(x;t)$ and mean energy density $\la\hat{H}(\bmv{r})\ra^t$ for 
quantum system with coupled states. We will suggest also one way of 
calculation of generalised memory functions in thermo field representation. 
It allows to analyse spectra of time correlation functions like 
``density-density'', ``current-current'' and
``energy-energy'' as well as generalised transport coefficients for quantum 
systems.

It is much sequential to describe investigations of kinetic and
hydrodynamic processes of a nuclear matter on the basis of quark-gluon
interaction. The quantum relativistic theory of kinetic and hydrodynamic
processes has its own problems and experiences its impetuous formation
\cite{c1,c2,c3,c4,c5,c6}. In the next papers we will apply our approach 
to describe kinetic and hydrodynamic processes of a quark-gluon plasma.

\end{document}